\documentclass[proof]{WileyASNA-v1}
\articletype{Original Article}%

\received{\today}
\revised{$ $Revision: 1.63 $ $}
\accepted{}

\usepackage{bm}
\graphicspath{{./fig/}{./png/}}

\newcommand{\EQ}{\begin{equation}}
\newcommand{\EN}{\end{equation}}
\newcommand{\EQA}{\begin{eqnarray}}
\newcommand{\ENA}{\end{eqnarray}}

\newcommand{\Eq}[1]{Equation~(\ref{#1})}

\newcommand{\Fig}[1]{Figure~\ref{#1}}

\newcommand{\Figs}[2]{Figures~\ref{#1} and \ref{#2}}

\newcommand{\Tab}[1]{Table~\ref{#1}}

\newcommand{\bra}[1]{\langle #1\rangle}

\newcommand{\meanPhi}{\overline{\Phi}}
{}
{}
{}

{}
{}
\newcommand{\meanEMF}{\overline{\mbox{\boldmath ${\cal E}$}}{}}{}
{}
{}
{}
{}
{}
\newcommand{\meanEE}{\overline{\mbox{\boldmath $E$}}{}}{}
{}
{}
{}
{}
{}
{}
{}
{}
\newcommand{\meanUU}{\overline{\bm{U}}}

{}

\newcommand{\meanA}{\overline{A}}
\newcommand{\meanB}{\overline{B}}

\newcommand{\meanJ}{\overline{J}}

{}

{}
{}
{}

\newcommand{\meanAA}{{\overline{\bm{A}}}}
\newcommand{\meanBB}{{\overline{\bm{B}}}}
\newcommand{\meanJJ}{{\overline{\bm{J}}}}

\newcommand{\xx}{\bm{x}}

\newcommand{\aaaa}{\bm{a}}
\newcommand{\jj}{\bm{j}}

\newcommand{\bb}{\bm{b}}
\newcommand{\BB}{\bm{B}}

\newcommand{\EE}{\bm{E}}
\newcommand{\JJ}{\bm{J}}
\newcommand{\oo}{\bm{\omega}}

\newcommand{\AAA}{\bm{A}}

\newcommand{\UU}{\bm{U}}

\newcommand{\uu}{\bm{u}}

\newcommand{\SSS}{\mbox{\boldmath $S$} {}}
\newcommand{\ee}{\mbox{\boldmath $e$} {}}

\newcommand{\ff}{\mbox{\boldmath $f$} {}}

\newcommand{\nab}{{\bm{\nabla}}}

\newcommand{\ii}{{\rm i}}

\newcommand{\DD}{{\rm D} {}}

\newcommand{\dd}{{\rm d} {}}

\def\Pm{\mbox{\rm Pr}_{\rm M}}
\def\Rm{\mbox{\rm Re}_{\rm M}}

\def\te{t_{\rm e}}
\def\cs{c_{\rm s}}

\def\kf{k_{\rm f}}

\def\Brms{B_{\rm rms}}

\def\urms{u_{\rm rms}}

\def\etat{\eta_{\rm t}}
\def\etatz{\eta_{\rm t0}}
\def\etatz{\eta_{\rm t0}}

\def\Beq{B_{\rm eq}}

\def\tautd{\tau_{\rm td}}
\def\tauto{\tau_{\rm to}}
\def\half{{\textstyle{1\over2}}}

\def\onethird{{\textstyle{1\over3}}}

\newcommand{\AU}{\,{\rm AU}}

\newcommand{\yapj}[3]{ #1, {ApJ,} {#2}, #3}

\newcommand{\yapjl}[3]{ #1, {ApJ,} {#2}, #3}

\newcommand{\yan}[3]{ #1, {Astron.\ Nachr.,} {#2}, #3}

\newcommand{\yana}[3]{ #1, {A\&A,} {#2}, #3}

\newcommand{\ygafd}[3]{ #1, {Geophys.\ Astrophys.\ Fluid Dyn.,} {#2}, #3}

\newcommand{\yjfm}[3]{ #1, {J.\ Fluid Mech.,} {#2}, #3}

\newcommand{\ypp}[3]{ #1, {Phys.\ Plasmas,} {#2}, #3}

\newcommand{\yprl}[3]{ #1, {Phys.\ Rev.\ Lett.,} {#2}, #3}

\newcommand{\ymn}[3]{ #1, {MNRAS,} {#2}, #3}
\newcommand{\ynat}[3]{ #1, {Nature,} {#2}, #3}

\newcommand{\ypre}[3]{ #1, {Phys.\ Rev.\ E,} {#2}, #3}

\newcommand{\yjour}[4]{ #1, {#2}, {#3}, #4}

\newcommand{\ybook}[3]{ #1, {#2} (#3)}

\def\apjs{ApJS}

\begin{document}

\title{Magnetic helicity and fluxes in an inhomogeneous $\alpha^2$ dynamo}

\author[1,2,3,4]{A. Brandenburg}

\authormark{A. Brandenburg}

\address[1]{\orgdiv{Nordita},
\orgname{KTH Royal Institute of Technology and Stockholm University},
\orgaddress{\state{Stockholm}, \country{Sweden}}}

\address[2]{\orgdiv{Department of Astronomy}, \orgname{Stockholm University},
\orgaddress{\state{Stockholm}, \country{Sweden}}}

\address[3]{\orgdiv{JILA and Laboratory for Atmospheric and Space Physics},
\orgname{University of Colorado},
\orgaddress{\state{Boulder}, \country{USA}}}

\address[4]{\orgdiv{McWilliams Center for Cosmology and Department of Physics},
\orgname{Carnegie Mellon University},
\orgaddress{\state{Pittsburgh}, \country{USA}}}

\corres{A. Brandenburg, Nordita,
KTH Royal Institute of Technology and Stockholm University,
10691 Stockholm, Sweden. \email{brandenb@nordita.org}}

\abstract{
Much work on turbulent three-dimensional dynamos has been done using
triply periodic domains, in which there are no magnetic helicity fluxes.
Here we present simulations where the turbulent intensity is still nearly
homogeneous, but now there is a perfect conductor boundary condition on
one end and a vertical field or pseudo-vacuum condition on the other.
This leads to migratory dynamo waves.
Good agreement with a corresponding analytically solvable $\alpha^2$
dynamo is found.
Magnetic helicity fluxes are studied in both types of models.
It is found that at moderate magnetic Reynolds numbers, most of the
magnetic helicity losses occur at large scales.
Whether this changes at even larger magnetic Reynolds numbers,
as required for alleviating the catastrophic dynamo quenching problem,
remains still unclear.
}


\fundingInfo{University of Colorado.
NSF Astronomy a Astrophysics Grants Program, 1615100.}

\maketitle

\section{Introduction}

Stars like the Sun possess large-scale magnetic fields that are
believed to be generated by an $\alpha$ effect, which amplifies
the magnetic field and sustains it against turbulent diffusive
decay \citep[see, e.g.,][]{RH04}.
Together with differential rotation, it can lead to cyclic magnetic
fields and equatorward migration \citep{Par55}.
The theoretical ``butterfly'' diagram of the mean toroidal magnetic
field versus time and latitude is similar to that observed \citep{SK69}.

The theory of the $\alpha$ effect of \cite{SKR66} is based on a
kinematic treatment, so the velocity field is assumed given.
Not only are the velocity fluctuations given, but one also assumes
that the magnetic fluctuations are entirely the result of tangling of
the large-scale magnetic field.
This is justified at small magnetic Reynolds numbers ($\Rm\ll1$,
corresponding to low electrical conductivity) or in cases when the
turbulence has a short correlation time.
The latter is an artificial construct, because real turbulence always
has a finite correlation time.
Therefore, the theory was always known to be problematic under
astrophysically relevant conditions when $\Rm\gg1$ and the correlation
time finite.

Significant progress was made by \cite{PFL76}, who extended the theory
of the $\alpha$ effect to a fully dynamical one, where $\alpha$ attains
a magnetic contribution corresponding to the swirl of the
magnetic field or, more precisely, the current helicity of the
small-scale magnetic field.
Their work also emphasized the importance of magnetic helicity
conservation, leading to an inverse cascade of magnetic helicity toward
large scales, but no explicit connection was drawn between the current
helicity of the small-scale field contributing to the $\alpha$ effect and
the large-scale magnetic field, as it builds up at the same time.

In an important paper by \cite{KR82}, the gap between the current helicity
contribution to the $\alpha$ effect and the actual large-scale magnetic
field was closed.
At that time, however, the focus was on the possibility
of chaos resulting from such a feedback \citep{Ruz81}, a topic that was
just introduced into solar physics by \cite{Tav78}.

Then, in the early 1990s, dynamo theory experienced a crisis with the work
of \cite{CV91} and \cite{VC92}, who found that turbulent diffusion and $\alpha$
effect are ``catastrophically'' quenched, i.e., they are quenched in an
$\Rm$-dependent fashion.
In a series of papers, \cite{GD94,GD95,GD96} developed an explanation in
terms of magnetic helicity conservation.
Using turbulent dynamo simulations in periodic domains, \cite{Bra01}
found large-scale magnetic fields in excess of the equipartition value,
but the ultimate nonlinear saturation phase lasted a microphysical
diffusion time, which would be very long in astrophysical applications.
It was only with the works of \cite{FB02}, \cite{BB02}, and \cite{Sub02}
that it became clear that the relevant theory describing such a quenching
is that of \cite{KR82}.

Unfortunately, the authors of this early work did not consider the
somewhat academic case of a homogeneous system, which was what
\cite{CH96} simulated.
They found a suppression of $\alpha$ proportional to
$1/(1+\Rm\meanBB^2/\Beq^2)$, suggesting that, in a saturated state,
$|\meanBB|$ would only be a fraction $\Rm^{-1/2}$ of the equipartition
field strength $\Beq$.
In an inhomogenous system, magnetic helicity fluxes are possible, and
those were already included in the theory of \cite{KR82}, but whether this
really alleviates the catastrophic quenching problem remains unclear
even today.
It is a hard problem, because even at magnetic Reynolds numbers of
1000, resistive contributions are still comparable to those of the magnetic
helicity flux \citep{HB10,MCCTB10,DSGB13}.

The basic idea of invoking magnetic helicity fluxes is the following.
As the $\alpha$ effect builds up a large-scale magnetic field, it also
builds up magnetic helicity associated with this large-scale field.
In the absence of magnetic helicity fluxes, small-scale magnetic
helicity of a sign opposite to that of the large-scale field must be produced
to obey total magnetic helicity conservation.
This is indeed what \cite{See96} and \cite{Ji99} found using independent
approaches.
The current helicity associated with small-scale magnetic helicity
quenches the dynamo prematurely.
Thus, to alleviate this quenching, small-scale magnetic helicity must be
removed preferentially.
This was demonstrated by \cite{BDS02} in what they called a ``vacuum
cleaner'' experiment, in which they removed all the small-scale magnetic
field from a simulated dynamo in regular intervals.
They found that this allows the field to saturate at a new level with
a stronger magnetic field.

The purpose of the present paper is to give an update on this situation
and to present new results of a model of an inhomogeneous dynamo that
may be suitable for addressing the problem of catastrophic quenching.
In particular, we will use an $\alpha^2$ dynamo that is made inhomogeneous
through the introduction of different boundary conditions on opposite
ends of the domain: a perfect connector (PC) boundary condition on the
lower end and a quasi-vacuum or vertical field (VF) condition on the upper.
This leads to an oscillatory $\alpha^2$ dynamo with dynamo waves
migrating away from the PC boundary to the VF boundary.
This is a system for which an analytic solution exists \citep{Bra17}.
Helicity fluxes of such a model were first studied using a mean field
model that incorporates magnetic helicity conservation
\citep[][hereafter BCC]{BCC09}.

The results of BCC remained puzzling in view of our understanding so far,
because they found that the flux of magnetic helicity at large and small
length scales was equally strong and of opposite sign.
This did not fit the expectation that there should be a preferential
loss of small-scale magnetic helicity.
Things became even more confusing when they found that the magnetic
helicity ejected into the halo outside the dynamo region has the wrong
sign; see the last panel of Fig.~7 of BCC.
Two years later, it was found that the magnetic helicity in the solar
wind at a distance of $1$--$5\AU$ from the Sun also has the wrong sign
\citep{BSBG11}.
Even today, there is no clear understanding of what this means in view of
the idea of alleviating catastrophic quenching by preferential small-scale
magnetic helicity losses.

The model of BCC has never been tested in a direct numerical simulation
(DNS).
This will be done in the present paper.
We begin by presenting the model and then report results from the
magnetic helicity fluxes in the resistive Weyl gauge \citep{CHBM11},
which is convenient for numerical purposes, although the results may a
priori be gauge-dependent.

\section{The model}

We use DNS similar to the homogeneous models of \cite{Bra01} and the
inhomogeneous ones of \cite{BD01}, except that here we have a PC boundary
condition at $z=0$ and a VF condition at $z=\pi/2$.
Such a choice was already adopted by BCC, and this size of the domain
was chosen because the eigenfunctions of the free decay problem are
quarter sine wave with wavenumber $k_1=1$.
In other words, the decay rate, $\etat k_1^2$, with a turbulent
diffusivity $\etat=1$ is then also unity.
We choose here a cubic domain, so its size is $(\pi/2)^3$.

We adopt an isothermal equation of state, so the pressure $p$ is
proportional to the density $\rho$ with $p=\rho\cs^2$, where $\cs$
is the isothermal sound speed.
We advance the magnetic vector potential $\AAA$ for the magnetic field
$\BB=\nab\times\AAA$, and solve the uncurled induction equation, the
momentum equation for the velocity $\UU$, and the continuity equation
for $\ln\rho$, so we have the following system of equations:
\EQ
{\partial\AAA\over\partial t}=\UU\times\BB+\eta\nabla^2\AAA,
\label{dAdt}
\EN
\EQ
\rho{\DD\UU\over\DD t}=\JJ\times\BB+\nab\cdot(2\nu\rho\SSS)
-\cs^2\nab\rho+\rho\ff,
\EN
\EQ
{\DD\ln\rho\over\DD t}=-\nab\cdot\UU,
\EN
where $\JJ=\nab\times\BB$ is the current density in units where
the vacuum permeability is unity,
${\sf S}_{ij}=\half(\partial_i u_j+\partial_j u_i)-\onethird
\delta_{ij}\nab\cdot\uu$ are the components of the traceless
rate-of-strain tensor, $\nu$ and $\eta$ are the kinematic
viscosity and magnetic diffusivity, respectively, and
$\ff$ is a monochromatic forcing function that
is $\delta$-correlated in time and fully helical so that
$\nab\times\ff\approx-\kf\ff$, i.e., the helicity is negative
and $\kf$ is the average forcing wavenumber.
We choose $\kf=16\,k_1$, which corresponds to 4 full wavelengths
across the $\pi/2$ domain.

In most of our simulations, the helicity of the forcing is
uniform, but in one case we adopt a modulation of the helicity
proportional to $\cos k_1 z$.
Thus, the helicity is then maximum at $z=0$, and goes to zero at
the boundary at $k_1z=\pi/2$.

Our simulations are characterized by the magnetic Reynolds and
magnetic Prandtl numbers,
\EQ
\Rm=\urms/\eta\kf,\quad
\Pm=\nu/\eta,
\EN
respectively, where $\urms=\bra{\uu^2}^{1/2}$ is the
root-mean squared (rms) velocity and angle brackets denote
volume averages.
By contrast, horizontal or $xy$ averages will be denoted
by overbars.
In this work we always take $\Pm=1$.

The magnetic field is often expressed in terms of the
equipartition field strength, $\Beq=\sqrt{\rho}\urms$.
We usually express time in turbulent--diffusive times,
$\tautd=(\etatz k_1^2)^{-1}$, where $\etatz=\urms/3\kf$
is an approximation to the turbulent magnetic diffusivity
\citep{SBS08}.
By comparison, the turnover time is $\tauto=(\urms\kf)^{-1}$,
which is shorter by the square of the scale separation or,
more precisely, $\tauto/\tautd=3(\kf/k_1)^2=768$.
Furthermore, the resistive time $\tau_\eta=(\eta k_1^2)^{-1}$
is longer than the turbulent--diffusive time by one third of
the magnetic Reynolds number, i.e., $\tau_\eta/\tautd=\Rm/3$.

\section{Results}

We consider three different Reynolds numbers using three different
resolutions.
Our model A with $\Rm=180$ and a resolution of $288^3$ mesh points
reaches full saturation and develops magnetic cycles with periods on
the scale of the turbulent--diffusive time, while models B and C
with higher $\Rm$ and higher resolution have not fully saturated.
In addition, we present a model with a gradient of the helicity and
hence a gradient of the underlying $\alpha$ effect (model~G), which
has otherwise similar parameters as model~A.
In fact, Run~G was the progenitor of Run~A.
All runs, along with their end time $\te$, expressed in different ways,
and their resolution are listed in \Tab{Tsum}.

\begin{table}[t!]\caption{
Parameters of the runs, their end times $\te$,
and the number of time steps $n_t$ in millions (M)
and the number of mesh points.
}\vspace{12pt}\centerline{\begin{tabular}{lccccccc}
Run & $\Rm$ & $\urms\kf\te$ & $\etatz k_1^2\te$ & $\eta k_1^2\te$ & $n_t$ & mesh \\
\hline
G & 170 &  7,000 &  9.1 & 0.16  & 3M & $288^3$ \\
A & 180 & 14,000 & 18.2 & 0.30  & 6M & $288^3$ \\
B & 370 &  2,700 &  3.5 & 0.028 &2.3M& $576^3$ \\
C & 750 &    300 &  0.4 & 0.0016&0.5M&$1152^3$ \\
\label{Tsum}\end{tabular}}\end{table}

\subsection{Magnetic field growth and saturation}

\begin{figure}[t!]\begin{center}
\includegraphics[width=\columnwidth]{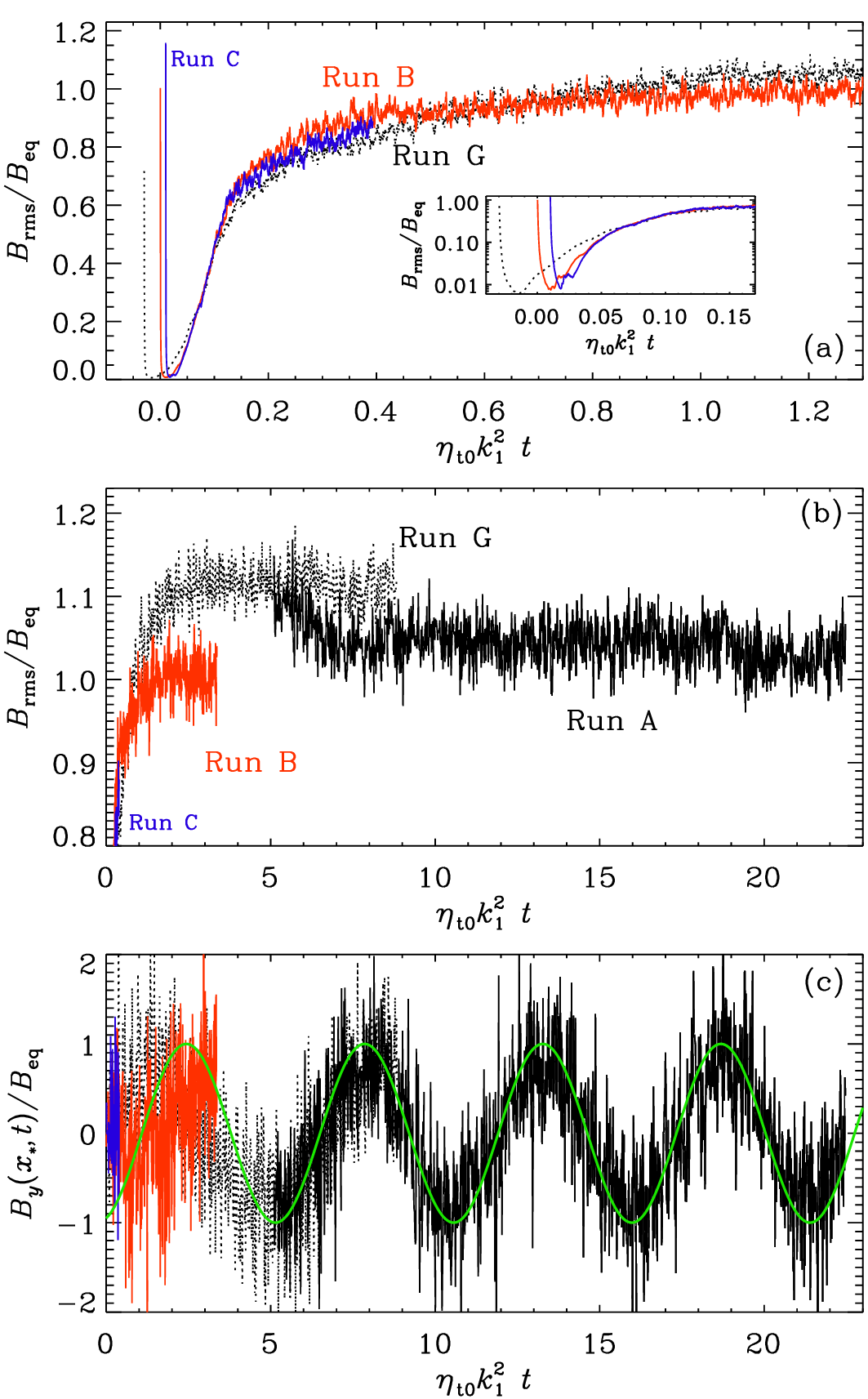}
\end{center}\caption[]{
Comparison of $\Brms/\Beq$ during the early saturation phase for
Runs~G, B, and C (a), the later development and restarting of Run~A
from Run~G at $\etatz k_1^2t\approx5$, and comparison with Runs~B
and C (b), and the establishment of magnetic cycles seen in the
magnetic field at one point $\xx=xx_\ast$ in the lower part of the
domain with frequency $\omega=1.16\etatz k_1^2$ (c).
}\label{pcomp}\end{figure}

We usually start with a random magnetic field that is $\delta$-correlated
in space.
This has the advantage that the dynamo quickly develops exponential
growth after the first 100 turnover times, or the first 0.2
turbulent--diffusive times; see \Fig{pcomp}.
In the present runs, the exponential growth phase is not well developed
because the initial field was already relatively strong.
In fact, it is initially close to equipartition, but, being
$\delta$-correlated in space, most of the field is in the diffusive
subrange and so the rms field drops by a factor of a hundred before the
dynamo sets in.

We can cautiously estimate the growth rate to be
$\lambda\approx0.09\urms\kf\approx70\etatz k_1^2$.
This is about three times faster than the typical growth rate of a mildly
supercritical small-scale dynamo, suggesting that it is a hybrid between
a small-scale and a large-scale dynamo.
The growth rates of such hybrid dynamos unifying small-scale and
large-scale dynamos was studied by \cite{SB14} and \cite{BSB16}
for a range of different values of $\Pm$.
They used, however, periodic boundary conditions, for which the time
for establishing a large-scale field is the resistive time \citep{Bra01}.

\begin{figure}[t!]\begin{center}
\includegraphics[width=\columnwidth]{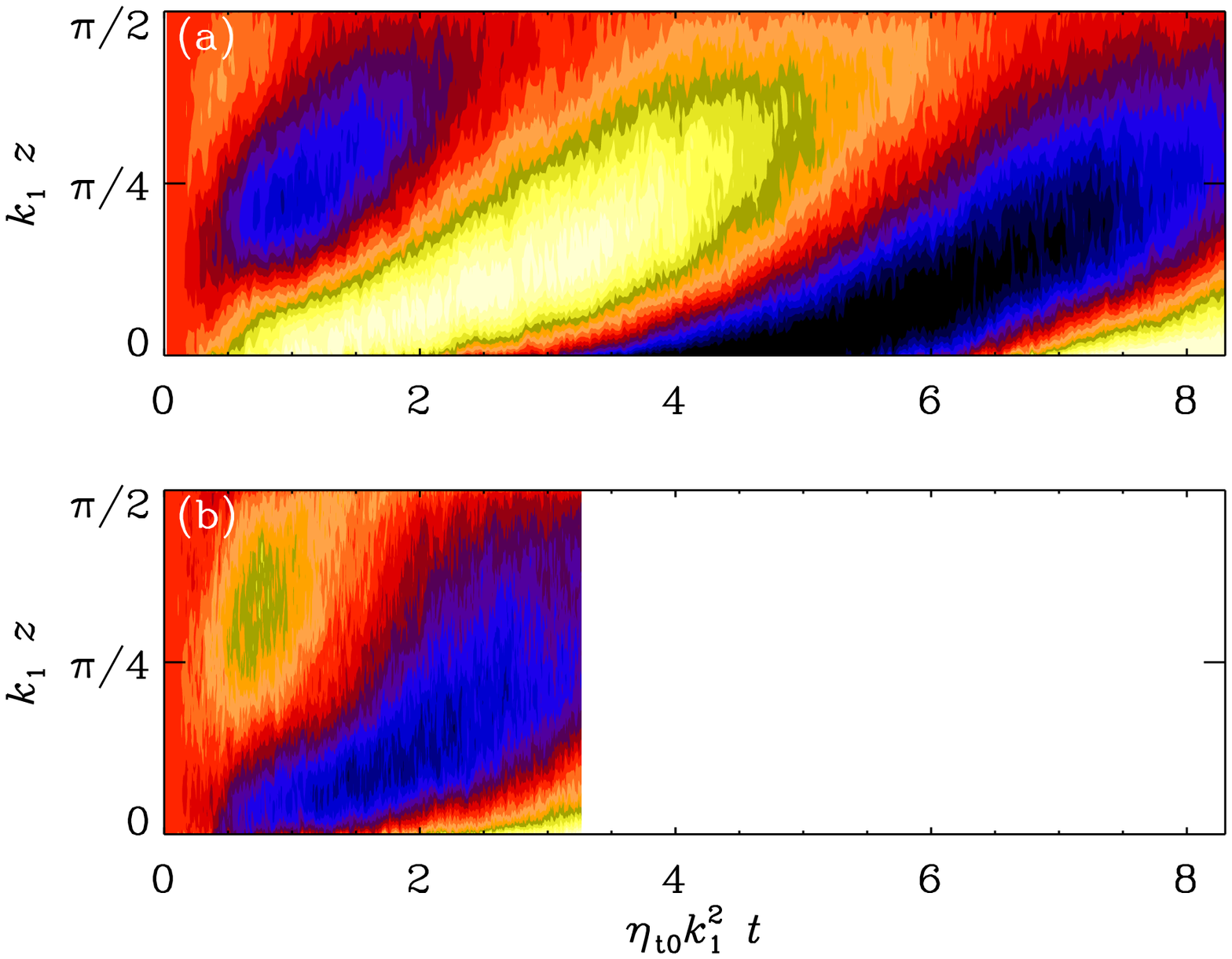}
\end{center}\caption[]{
Butterfly diagrams of $\meanB_y(z,t)$ for Runs~G (a) and B (b) during
the first eight turbulent--diffusive times.
Note that in both cases a large-scale field is clearly visible after
a fraction of the turbulent--diffusive time.
}\label{ppbutter_comp}\end{figure}

In addition to the total (small-scale plus large-scale) magnetic field
reaching rapid saturation without prolonged nonlinear saturation phase,
as in \cite{Bra01} and \cite{CB13}, also the space-time properties of
the large-scale magnetic field are found to develop quickly.
This is shown in \Fig{ppbutter_comp}, where we compare butterfly diagrams
for Run~G with $\Rm=170$ and Run~B with $\Rm=370$.
In both cases the large-scale field is clearly visible even after just
one turbulent--diffusive time, or even earlier.
Nevertheless, the rms magnetic field is still growing, as can be seen
from \Fig{pcomp}.
Unfortunately, such runs are getting computationally rather costly
and even after 2.3 million time steps they have not run for long
enough to see a well developed cycle.

\begin{figure*}[t!]\begin{center}
\includegraphics[width=\textwidth]{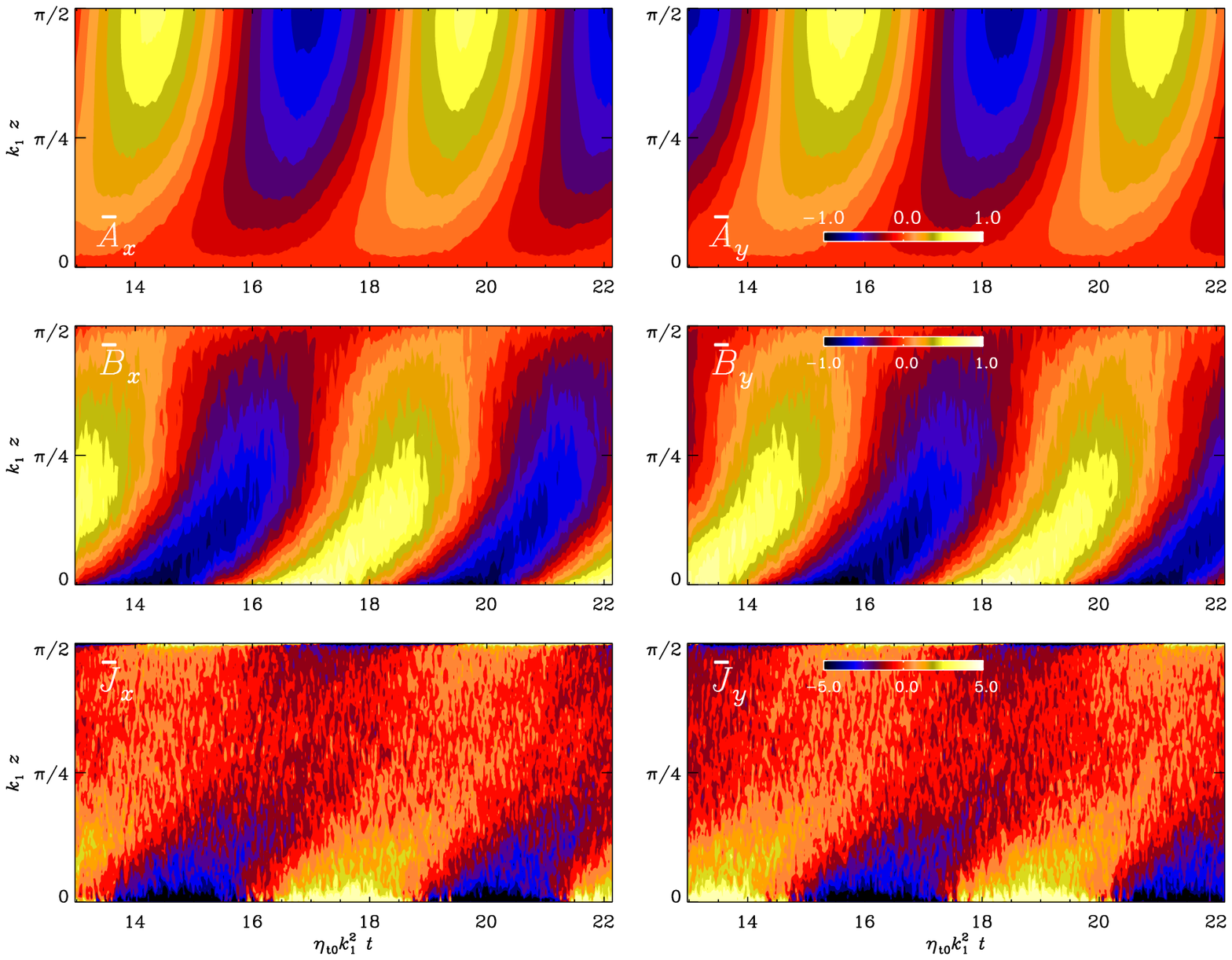}
\end{center}\caption[]{
Butterfly diagrams of $\meanA_x$, $\meanA_y$, $\meanB_x$, $\meanB_y$,
$\meanJ_x$, and $\meanJ_y$ for Run~A, covering almost two cycles
near the end of the run.
}\label{ppbutter_M288h_cont}\end{figure*}

\subsection{Butterfly diagrams for other fields}

In \Fig{ppbutter_M288h_cont}, we present butterfly diagrams of $\meanA_x$,
$\meanA_y$, $\meanB_x$, $\meanB_y$, $\meanJ_x$, and $\meanJ_y$.
All panels show clear signs of the magnetic cycle.
However, $\meanJ_x$ and $\meanJ_y$ are rather ``noisy'', which is
connected with the fact that they are related to higher derivatives
of $\AAA$ and $\BB$.

Note that, because of $\nab\cdot\meanBB=\nab\cdot\meanJJ=0$, and because
there is no mean flux, we have $\meanB_z=\meanJ_z=0$.
In our case $\nab\cdot\meanAA$ does not vanish, so $\meanA_z$ does not
vanish either, it does not display any cyclic variations and is not of
physical interest because it does not contribute to $\meanBB$.
We return to the discussion of $\nab\cdot\AAA$ in connection with magnetic
helicity fluxes, but even then it will turn out to be unimportant.

\begin{figure*}[t!]\begin{center}
\includegraphics[width=\textwidth]{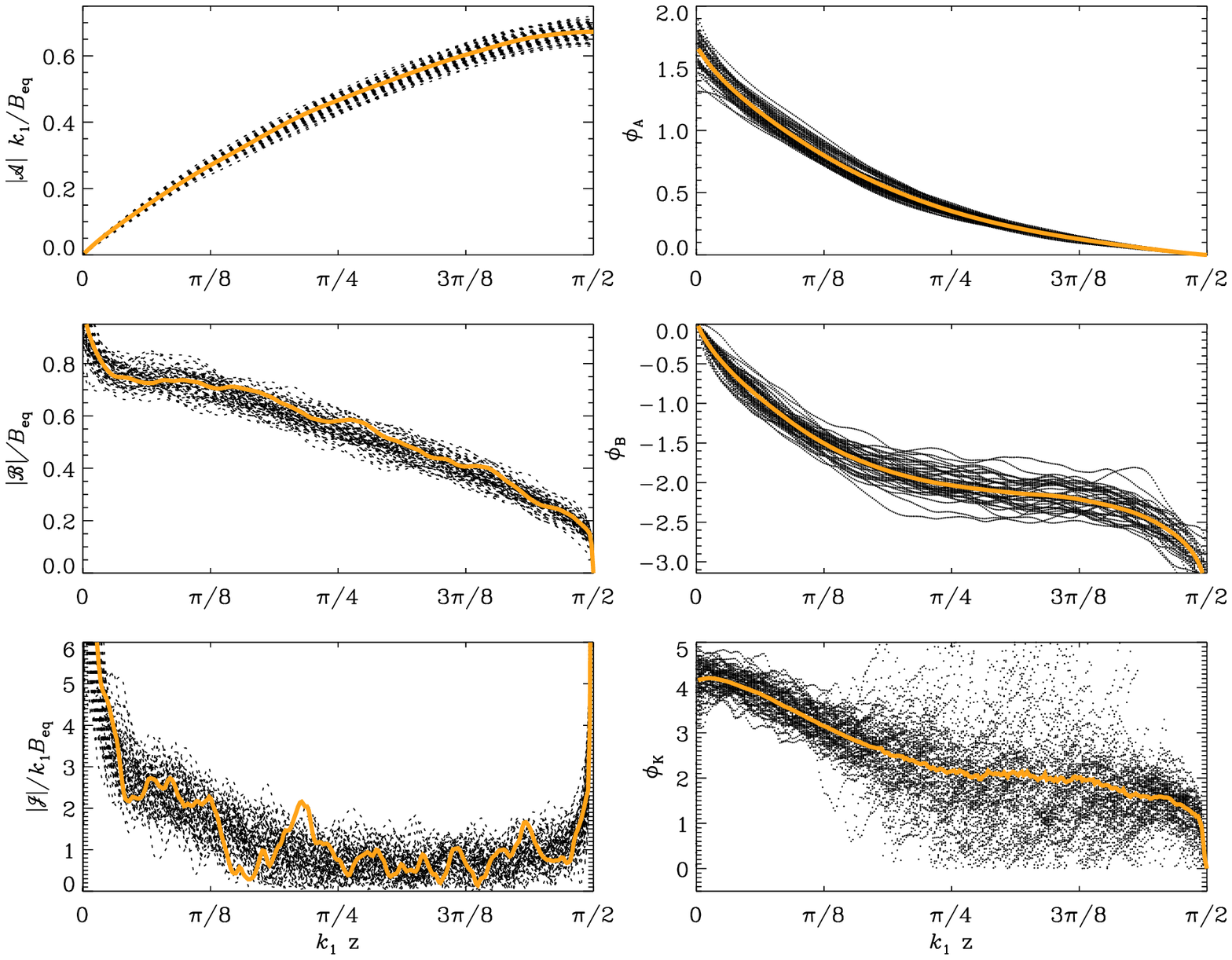}
\end{center}\caption[]{
Moduli and phases of ${\cal A}$, ${\cal B}$, ${\cal J}$
for Run~A as a function of $z$.
The fat orange lines denote the temporal averages.
}\label{ppbutter_complex_M288h_cont}\end{figure*}

The large-scale magnetic field varies in the $z$ direction and is roughly
described by what is expected based on the analytic mean-field theory.
It is useful to express the mean magnetic fields in complex notation
as \citep{Bra17}
\EQA
&{\cal A}\equiv\meanA_x+\ii\meanA_y=r_A(z)\,e^{\ii\phi_A(z)-\ii\omega t},\\
\ii\partial{\cal A}=&{\cal B}\equiv\meanB_x+\ii\meanB_y=r_B(z)\,e^{\ii\phi_B(z)-\ii\omega t},\\
\ii\partial{\cal B}=&{\cal J}\equiv\meanJ_x+\ii\meanJ_y=r_J(z)\,e^{\ii\phi_J(z)-\ii\omega t}.
\ENA
Thus, at each time, the complex quantities ${\cal A}$, ${\cal B}$,
and ${\cal J}$ experience a certain phase shift.
However, since we have a PC boundary condition with ${\cal A}=0$ on $z=0$,
we can subtract the phase at $z=0$ at each time and overplot the results
in a single plot.
Likewise for ${\cal B}$, which vanishes on the VF condition at
$k_1 z=\pi/2$, we subtract here the phase at $k_1 z=\pi/2$ and overplot.
For ${\cal J}$, no such condition exists, but we still get a reasonable
result by subtracting the phase at $z=0$ at each time.
For the moduli of ${\cal A}$, ${\cal B}$, and ${\cal J}$, no complications
arise.
The results for the nonlinearly saturated fields are plotted in
\Fig{ppbutter_complex_M288h_cont}, normalized by $\Beq/k_1$, $\Beq$,
and $\Beq\,k_1$, respectively.

Again, the current density appears more noisy than the quantities
in any of the other plots.
Somewhat more surprising this time is the fact that $|{\cal J}|$
is mostly flat, except in the proximity of both boundaries.
This could possibly hint at numerical artifacts related to the
fact that we have the PC and VF boundary conditions as symmetry
conditions, which also affect second and higher derivatives
in unwanted ways.
As shown in \cite{Bra17}, such an approach still gives correct
solutions, but they converge more slowly.

\begin{figure*}[t!]\begin{center}
\includegraphics[width=\textwidth]{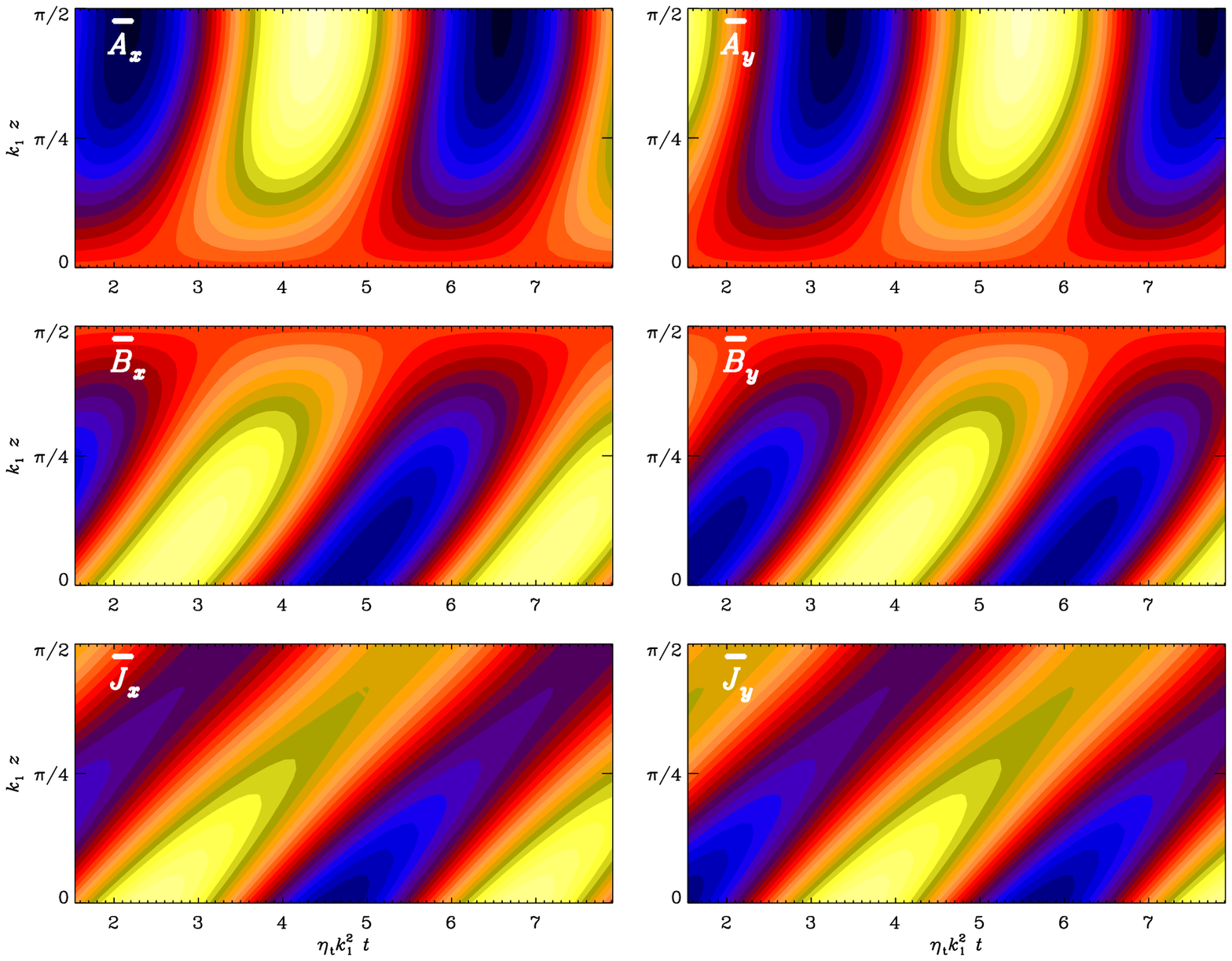}
\end{center}\caption[]{
Butterfly diagrams of $\meanA_x$, $\meanA_y$, $\meanB_x$, $\meanB_y$,
$\meanJ_x$, and $\meanJ_y$ for the $\alpha^2$ mean-field dynamo.
The color bars are similar to those in \Fig{ppbutter_M288h_cont},
except that the amplitude is undetermined in linear theory.
}\label{pcontour3}\end{figure*}

\begin{figure*}[t!]\begin{center}
\includegraphics[width=\textwidth]{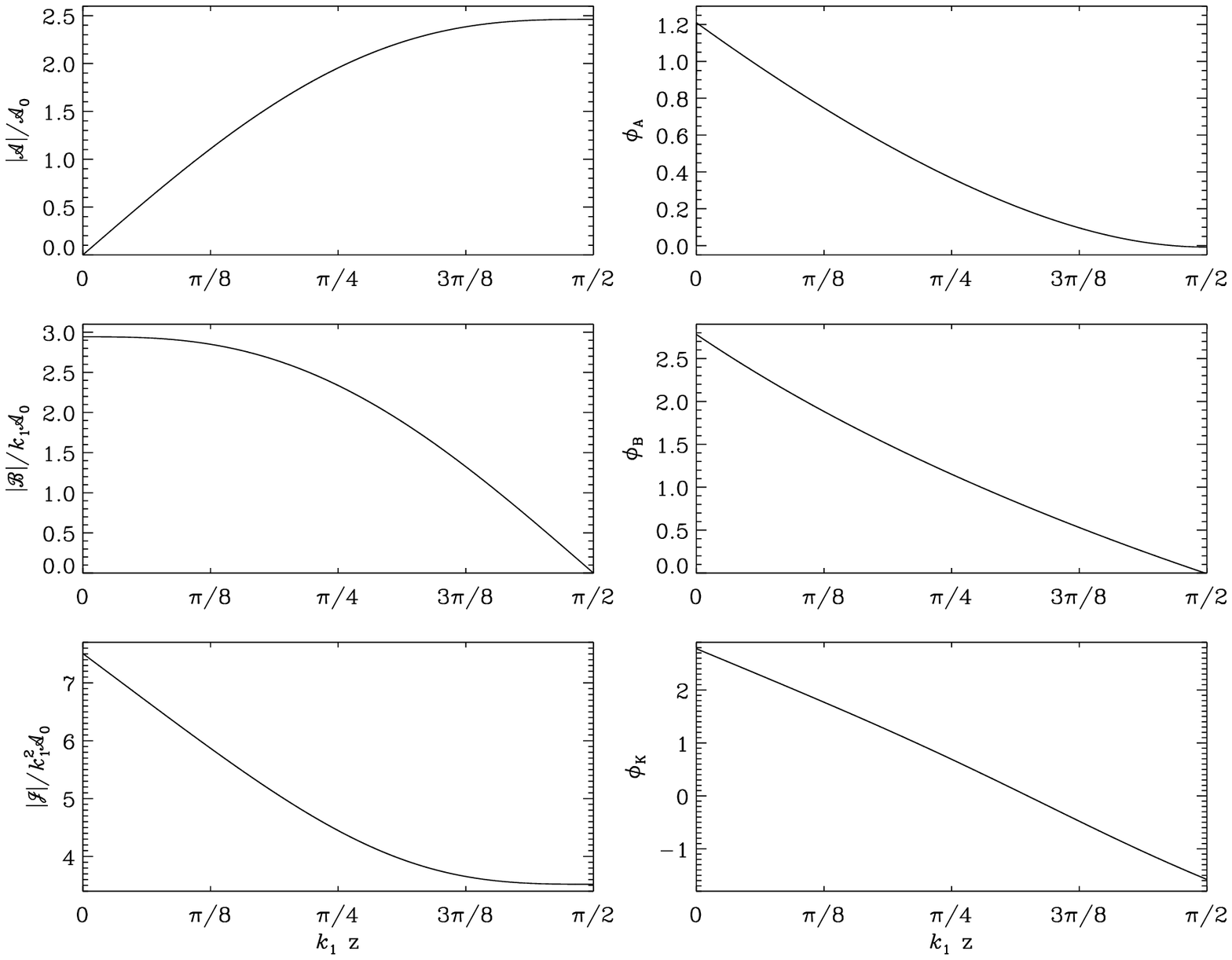}
\end{center}\caption[]{
Moduli and phases of ${\cal A}$, ${\cal B}$, ${\cal J}$
for the $\alpha^2$ of \cite{Bra17} as a function of $z$.
}\label{pplot2_ABJ}\end{figure*}

\subsection{Comparison with the analytic $\alpha^2$ dynamo}

The solution of the $\alpha^2$ dynamo with constant $\alpha$ and turbulent
magnetic diffusivity $\etat$, and with PC boundary conditions on $z=0$
and a VF condition on $z=\pi/2$ reads \citep{Bra17}
\EQ
{\cal A}(z,t)={\cal A}_0\,\left(e^{\ii k_+z}-e^{\ii k_-z}\right)\,
e^{-\ii\omega t},
\EN
where ${\cal A}_0$ is an amplitude factor, which is undefined
in linear theory.
Furthermore,
\begin{eqnarray}
k_+/k_1\approx0.10161896-0.51915398\,\ii,\\
k_-/k_1\approx-2.6522693+0.51915398\,\ii,
\end{eqnarray}
are complex wavenumbers, and
\begin{equation}
\omega/\etat k_1^2\approx-1.4296921
\label{eigenval}
\end{equation}
is the frequency for the marginally excited dynamo with the critical
value $\alpha\approx2.5506504\etat k_1$.

In \Fig{pcontour3}, we plot butterfly diagrams of the real and imaginary
parts of ${\cal A}(z,t)$, ${\cal B}(z,t)$, and ${\cal J}(z,t)$ for this
analytic solution.
In comparison with \Fig{ppbutter_M288h_cont}, the main difference is
that in the simulation the pattern develops a large migration speed
at large values of $z$, i.e., the butterfly wings are nearly vertical.
This is not seen in the analytic model.
Apart from this, however, the cycle period is similar:
$\omega=1.16\,\etat k_1^2$ in the simulation compared with 
$1.43\,\etat k_1^2$ in the analytic model, which is about 23\% larger.

In \Fig{pplot2_ABJ} we show the absolute values and phases of
${\cal A}$, ${\cal B}$, ${\cal J}$ for this $\alpha^2$ dynamo as a
function of $z$.
All these dependencies are similar to those in the three-dimensional
simulations; see \Fig{ppbutter_complex_M288h_cont}.
However, there are also systematic differences, but it is unclear to
what extent those are related to the fact that our simulations are
nonlinear, and that the effective $\alpha$ and $\etat$ acting in this
three-dimensional simulations may not be constant in space.
There may be many other such reasons for the disagreement.

To find out how supercritical the mean-field dynamo in the simulation is,
we estimate the modulus of $\alpha$ as $\alpha_0=\urms/3$ \citep{SBS08}
and compute the dynamo number as $C_\alpha=\alpha_0/\etatz k_1=\kf/k_1=4$.
Thus, the dynamo is less than 1.6 times supercritical, which is not much.
It would be interesting to compare with solutions that are either more
supercritical or less supercritical, but this will not be done in the
present work.
Instead, in the rest of this work, we focus on magnetic helicity fluxes.

\begin{figure}[t!]\begin{center}
\includegraphics[width=\columnwidth]{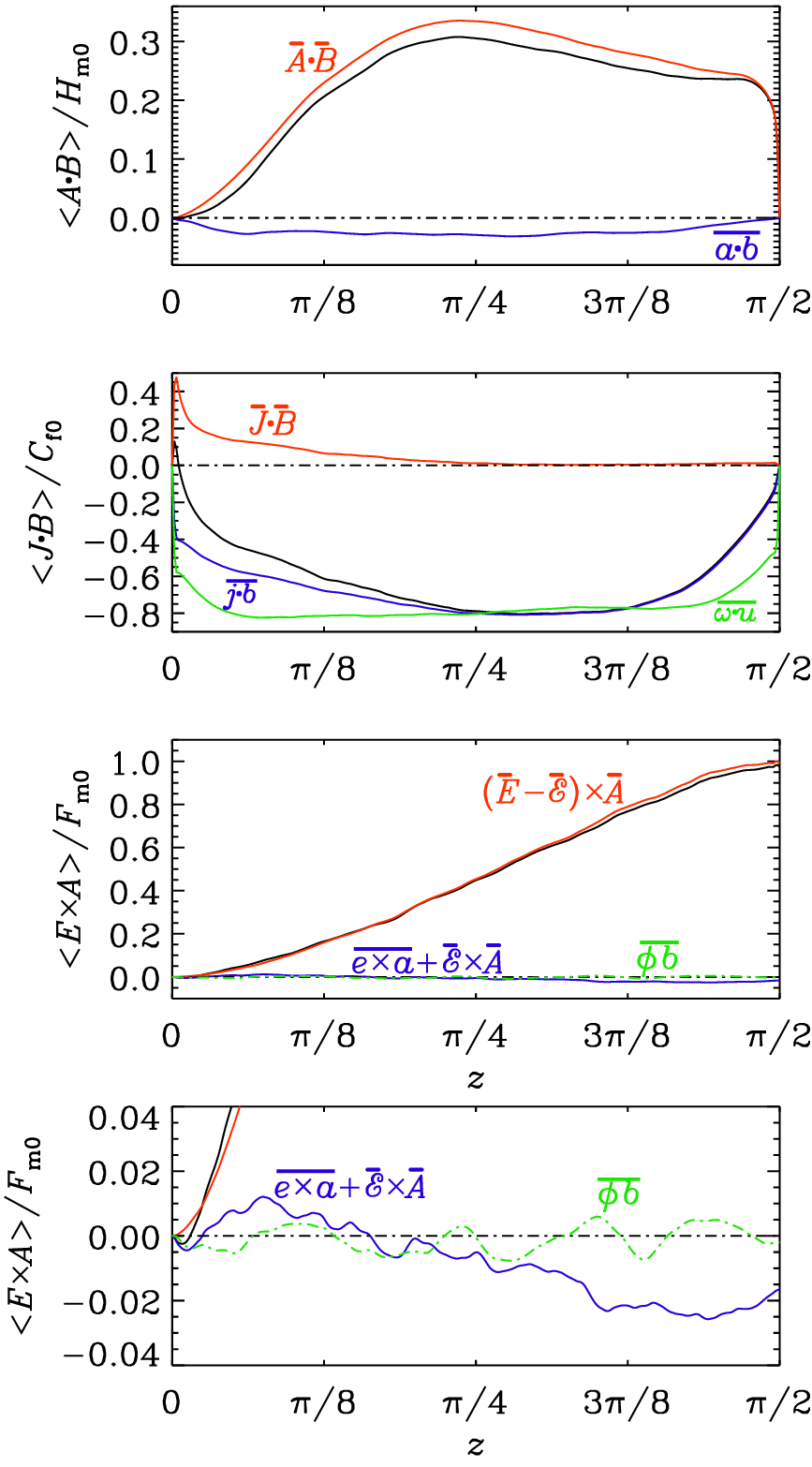}
\end{center}\caption[]{
Magnetic helicity, current helicity, and magnetic helicity fluxes
for Run~A with $\Rm=180$.
The black lines give the total value, red lines give the contribution
from the large-scale fields, and blue lines the contribution from
small-scale fields.
In the second panel, the kinetic helicity is shown in green
and is found to be of similar magnitude as the current helicity
of the small-scale field.
The last two panels are similar except that a smaller range
in $\overline{\EE\times\AAA}$ near zero is shown.
Here the green line denotes $\overline{\phi\bb}$, which is seen
to fluctuate around zero.
In all cases, only the $z$ components of the fluxes are plotted.
}\label{pphelflux_M288h_cont2}\end{figure}

\begin{figure}[t!]\begin{center}
\includegraphics[width=\columnwidth]{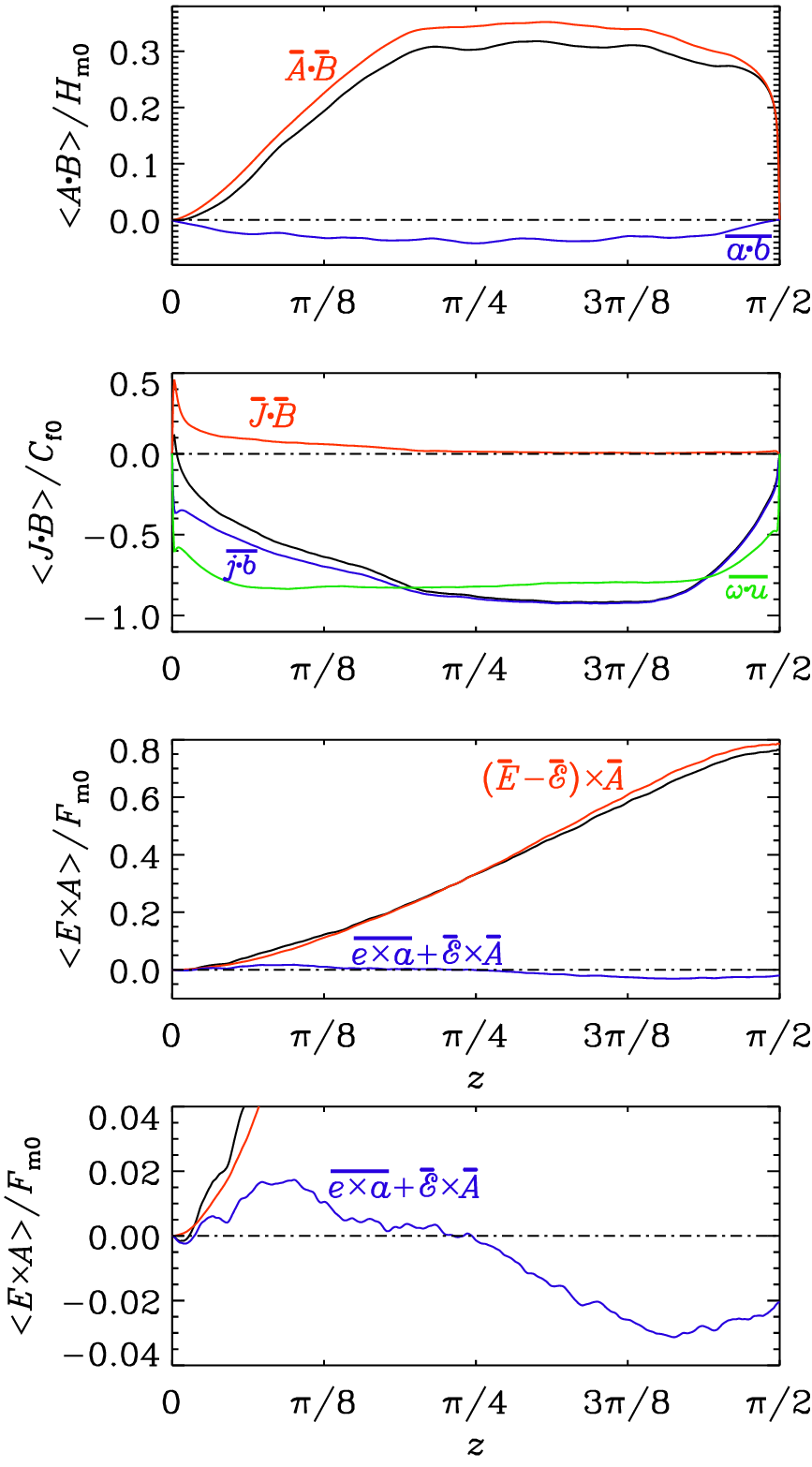}
\end{center}\caption[]{
Similar to \Fig{pphelflux_M288h_cont2}, but for Run~B with $\Rm=370$.
}\label{pphelflux_M576a}\end{figure}

\begin{figure}[t!]\begin{center}
\includegraphics[width=\columnwidth]{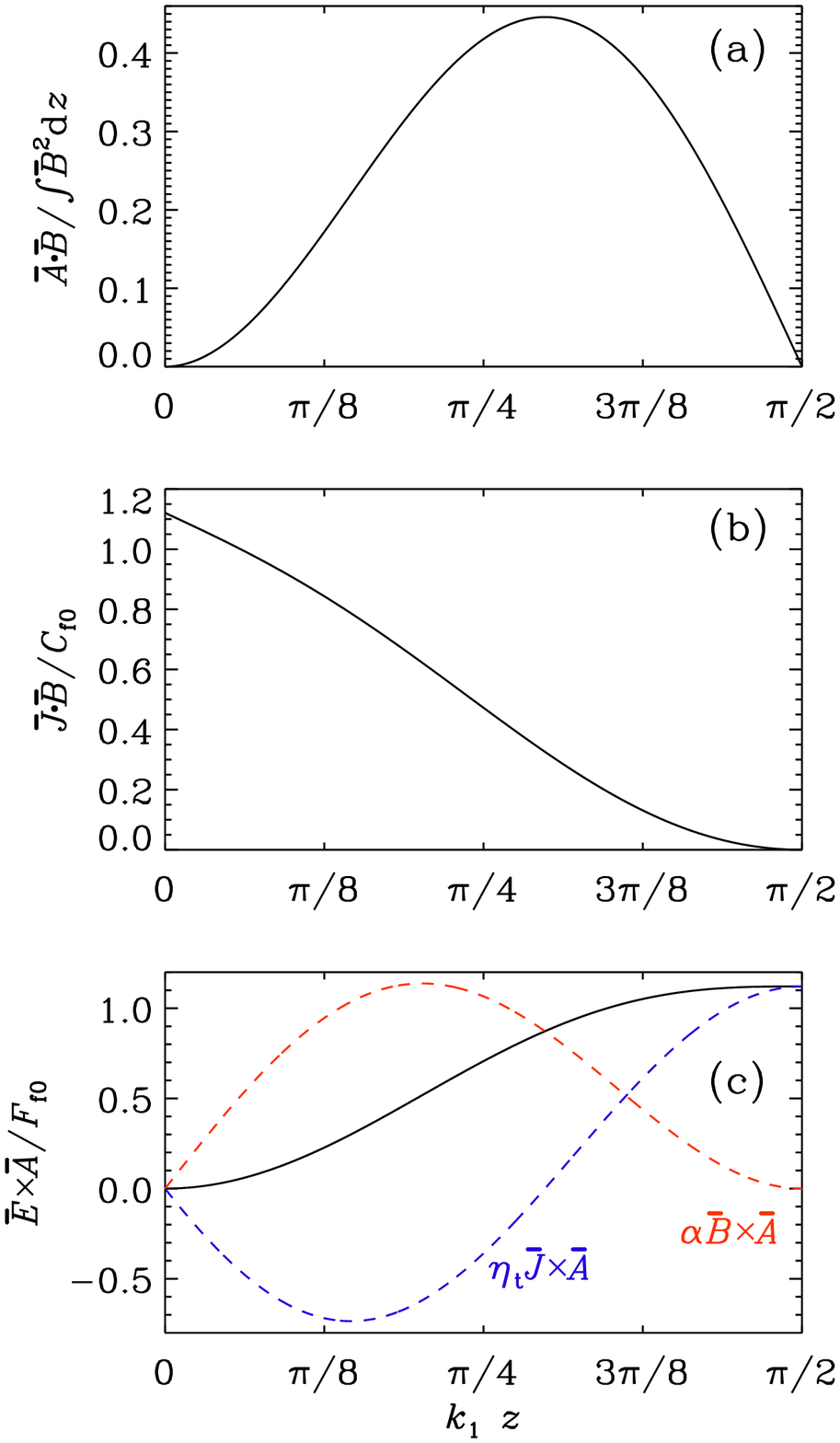}
\end{center}\caption[]{
Magnetic and current helicity profiles as well as
magnetic helicity fluxes in the mean-field $\alpha^2$ dynamo.
}\label{pplot3}\end{figure}

\begin{figure}[t!]\begin{center}
\includegraphics[width=\columnwidth]{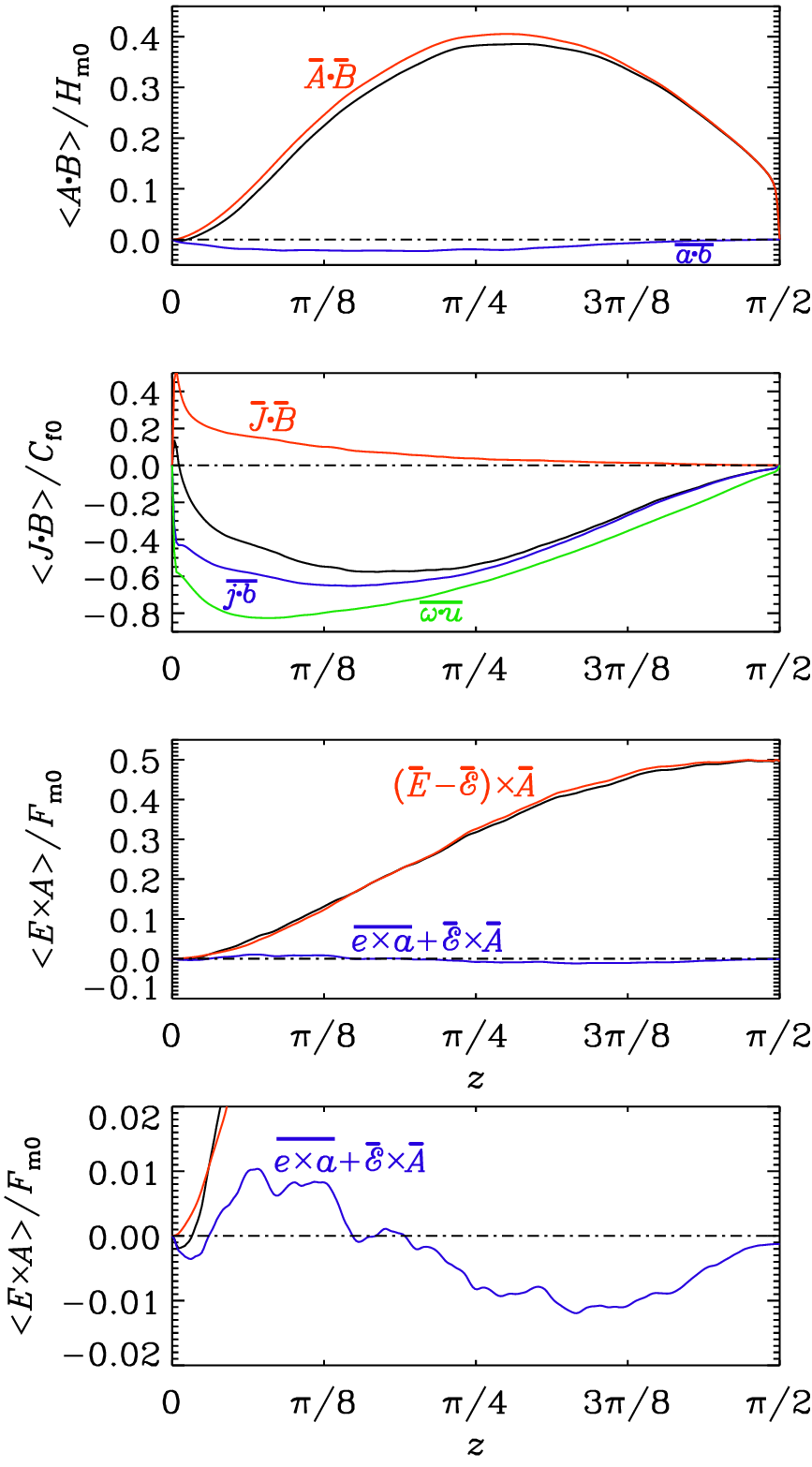}
\end{center}\caption[]{
Similar to \Fig{pphelflux_M288h_cont2}, but for Run~G with $\Rm=170$.
}\label{pphelflux_M288g}\end{figure}

\subsection{Magnetic helicity balance}

We now discuss the magnetic helicity equation, which is obtained by
dotting \Eq{dAdt} with $\BB$ and adding the curl of \Eq{dAdt} dotted
with $\AAA$.
The result is
\EQ
{\partial\over\partial t}\overline{\AAA\cdot\BB}=
-2\eta\overline{\JJ\cdot\BB}-{\partial\over\partial z}
\left(\overline{\EE\times\AAA}+\overline{\Phi\BB}\right),
\label{dABmdt}
\EN
where $\EE=\eta\JJ-\UU\times\BB$ is the electric field and
$\Phi=-\eta\nab\cdot\AAA$ is a scalar potential that results here from
the facts that $\eta$ is constant and the diffusion term on the right-hand
side in \Eq{dAdt} can be written as $\nabla^2\AAA=-\JJ+\nab\nab\cdot\AAA$.
Next, we consider the evolution of the magnetic helicity of the mean
field, $\meanAA\cdot\meanBB$, and subtract it from \Eq{dABmdt} to obtain
the evolution equation of the magnetic helicity of the fluctuating fields,
$\aaaa=\AAA-\meanAA$ and $\bb=\BB-\meanBB$, using
\EQ
\overline{\aaaa\cdot\bb}=\overline{\AAA\cdot\BB}-\meanAA\cdot\meanBB,
\EN
where we have made use of the Reynolds rules for horizontal averages.
The evolution equation for $\meanAA$ can be written in the form
\EQ
{\partial\meanAA\over\partial t}=\meanUU\times\meanBB+\meanEMF
-\eta\meanJJ-\nab\meanPhi,
\EN
where $\meanPhi=-\eta\nab\cdot\meanAA=-\eta\partial\meanA_z/\partial z$,
and the $\eta\meanJJ$ term describes the microphysical diffusion of
the mean field, which is usually small in comparison with turbulent
magnetic diffusion.
The traditional $\alpha$ effect and turbulent diffusion are all
modeled through the $\meanEMF=\overline{\uu\times\bb}$ term as
$\meanEMF=\alpha\meanBB-\etat\meanJJ$, but no
such specification needs to be made at this point.
Thus, we have
\EQ
{\partial\over\partial t}(\meanAA\cdot\meanBB)=
2\meanEMF\cdot\meanBB-2\eta\meanJJ\cdot\meanBB
-{\partial\over\partial z}
\left(\meanEE\times\meanAA+\meanPhi\,\meanBB
-\meanEMF\times\meanAA\right),
\label{dAmBmdt}
\EN
where $\meanEE\equiv\eta\meanJJ-\meanUU\times\meanBB$ is defined as the electric
field resulting from the mean fields, but without the contribution from the
fluctuations that are already included in the $\meanEMF$ term.
By subtracting \Eq{dAmBmdt} from \Eq{dABmdt}, we obtain
\EQ
{\partial\over\partial t}\overline{\aaaa\cdot\bb}=
-2\meanEMF\cdot\meanBB
-2\eta\overline{\jj\cdot\bb}-{\partial\over\partial z}
\left(\overline{\ee\times\aaaa}+\overline{\phi\bb}
+\meanEMF\times\meanAA\right),
\label{dafbfdt}
\EN
where
\EQ
\overline{\jj\cdot\bb}=\overline{\JJ\cdot\BB}-\meanJJ\cdot\meanBB
\EN
is the current helicity of the small-scale field,
\EQ
\overline{\ee\times\aaaa}
=\overline{\EE\times\AAA}-\meanEE\times\meanAA
\EN
is the magnetic helicity flux of the small-scale field, and
\EQ
\overline{\phi\bb}=\overline{\Phi\BB}-\meanPhi\,\meanBB
\EN
is a contribution to the magnetic helicity flux that results from
the particular gauge of the small-scale field.

In \Figs{pphelflux_M288h_cont2}{pphelflux_M576a}, we compare the profiles
of magnetic helicity, current helicity, and the magnetic helicity fluxes
for Runs~A and B with $\Rm=180$ and $370$, respectively.
For normalization purposes, we have defined
\EQA
&H_{\rm m0}=\int_0^{\pi/2}\meanBB^2\,\dd z,\\
&C_{\rm f0}=\kf\Beq^2,\\
&F_{\rm m0}=\etatz k_1^2\int_0^{\pi/2}\meanBB^2\,\dd z.\\
\ENA
Not surprisingly, the largest contribution to the magnetic helicity
density comes from the large-scale field.
This is also reasonably well reproduced by the mean-field model; see
\Fig{pplot3}.

Again, not surprisingly, the current helicity is dominated by the
small-scale parts with $\overline{\jj\cdot\bb}/\kf\Beq^2$ being
reasonably close to $-1$.
The current helicity of the small-scale field is also close to the
kinetic helicity density, similarly to what was found for perfectly
homogeneous dynamos in periodic domains \citep{Bra01}.
More surprising is the fact that most of the magnetic helicity flux
comes from the large-scale magnetic field, and very little from the
small-scale field.
However, it may be interesting that the contribution from the small-scale
field is similar for Runs~A and B, i.e.,
$\overline{\EE\times\AAA}/F_{\rm m0}\approx-0.03$ in both cases,
and perhaps even slightly larger for Run~B.
It will be interesting to evaluate these terms at even higher resolution
and for higher magnetic Reynolds numbers.
It is conceivable that the fluxes from the large-scale field continue
to decline, but that those of the small-scale field remain constant or
increase and might eventually be equal to the contributions from the
small-scale fields.

\subsection{Dependence on $\Rm$}

To study the dependence of the solutions on $\Rm$, it is useful to
compute the energy contained in the mean field, which is essentially
the same as our normalization constant $H_{\rm m0}$.
In \Tab{THm0} we list those values for all four runs and normalize by
$\Beq^2/k_1$, which itself is almost the same in all four cases.

\begin{table}[b!]\caption{
Comparison of $H_{\rm m0}$, normalized by $\Beq^2/k_1$, for all of
our runs.
For Run~C, only a lower limit is given, because it is still too short.
}\vspace{12pt}\centerline{\begin{tabular}{lccccccc}
Run & $\Rm$ & mesh & $H_{\rm m0} \, k_1/\Beq^2$ \\
\hline
G & 170 &  $288^3$ & 0.60 \\
A & 180 &  $288^3$ & 0.51 \\
B & 370 &  $576^3$ & 0.36 \\
C & 750 & $1152^3$ & $>0.02$ \\
\label{THm0}\end{tabular}}\end{table}

We see that there is a systematic decline in $H_{\rm m0}\,k_1/\Beq^2$
as $\Rm$ increases.
Thus, these models do not yet appear to be in the asymptotic regime.
Investigating these values for longer simulations and at larger values of
$\Rm$ will be important and is also needed for calculating reliable
error bars.

\subsection{Comparison with Run~G}

Run~G had a vertical gradient in the helicity density, while the turbulent
intensity was approximately independent of $z$.
It is possible that such a gradient in the kinetic helicity,
$\overline{\oo\cdot\uu}$, where $\oo=\nab\times\uu$ is the vorticity,
would produce a similar gradient both in $\overline{\jj\cdot\bb}$ and
in $\overline{\aaaa\cdot\bb}$, and thereby a magnetic helicity flux
proportional to $-\nab\overline{\aaaa\cdot\bb}$, as proposed by \cite{HB11}.
This does not seem to be the case and the magnetic helicity flux for the
small-scale field does even go to zero near the upper boundary; see the
last panel of \Fig{pphelflux_M288g}.
Some other features in the nature of the small-scale turbulence may be
needed that have not yet been identified.

\section{Conclusions}

The present work has shown that turbulent dynamos with homogeneous
helical forcing, but different boundary conditions on the lower and
upper boundaries, lead to dynamo waves and magnetic helicity fluxes --
similarly to what is expected based on a mean-field model.
Both mean-field models and simulations are similar in many respects
and agree qualitatively within 23\% in terms of the cycle period, but
there are differences in the shape of the magnetic field and current
density profiles.

The magnetic helicity profiles are strongly dominated by the large-scale
magnetic field.
This is somewhat disappointing in the sense that, to alleviate the
catastrophic quenching problem discussed in the introduction, we expect
the magnetic helicity fluxes to be dominated by small-scale contributions.
On the other hand, it may be this magnetic helicity of the large-scale
field that contributes to the mysterious sign reversal found originally
in the solar wind \citep{BSBG11}.
If this is correct, it may indicate that even in the Sun, the catastrophic
quenching problem has not quite gone away yet.

Comparing with Run~G, which has a gradient in the kinetic helicity
profile, we saw that it does not lead to major differences, suggesting
that small-scale magnetic helicity transport from a downward gradient
of magnetic helicity density is not very efficient in this model.
It may therefore be useful to reconsider some of the earlier setups that
have been studied to measure magnetic helicity fluxes in dynamo simulations.
\cite{MCCTB10} considered a model between perfectly conducting boundaries,
while \cite{HB10} considered a model with a poorly conducting halo.
Finally, \cite{DSGB13} studied a model where both advection from a
wind and a downward gradient of magnetic helicity contributed
to the flux of magnetic helicity.
Comparative studies of the magnetic helicity fluxes from the small-scale
magnetic field will be the subject of a separate publication.

\section*{\scriptsize{Acknowledgements}}
I thank Gherardo Valori and Etienne Pariat for organizing the Magnetic
Helicity in Astrophysical Plasmas Team at the International Space Science
Institute in Bern, where this work began.
This research was supported in part by the Astronomy and Astrophysics
Grants Program of the National Science Foundation (grant 1615100),
and the University of Colorado through
its support of the George Ellery Hale visiting faculty appointment.
We acknowledge the allocation of computing resources provided by the
Swedish National Allocations Committee at the Center for Parallel
Computers at the Royal Institute of Technology in Stockholm.


\vfill\bigskip\noindent\tiny\begin{verbatim}
$Header: /var/cvs/brandenb/tex/mhd/helflux_a2dynamo/paper.tex,v 1.63 2019/02/04 10:27:00 brandenb Exp $
\end{verbatim}

\end{document}